\documentclass[preprint]{nature}

\usepackage{graphicx}

\countdef\preprint=0 \countdef\tube=1
\ifnum\tube<10

\else

\fi

\begin{document}
\title{Graphite intercalation compound KC$_8$ revisited: a key to graphene}
\maketitle


\author{A.~Gr\"{u}neis$^{1}$,
C.~Attaccalite$^{2}$,  A.~Rubio$^{2}$,  D.~Vyalikh$^3$,
S.L.~Molodtsov$^3$, J.~Fink$^4$, R.~Follath$^4$, W.~Eberhardt$^4$, B.~B\"uchner$^1$, T.~Pichler$^5$}

\begin{affiliations}
\item IFW Dresden, P.O. Box 270116, D-01171 Dresden, Germany \item
Nano-Bio Spectroscopy Group and European Theoretical Spectroscopy
Facility (ETSF), Departamento de Fisica de Materiales, Unidad de
Materiales Centro Mixto CSIC-UPV/EHU, Universidad del Pais Vasco,
Avd. Tolosa 72, E-20018 Donostia, Spain \item
Institut f\"ur Festk\"orperphysik, TU Dresden, Mommsenstrasse 13,
D-01069 Dresden, Germany \item BESSY II, Albert-Einstein-Str. 15,
D-12489 Berlin, Germany \item Faculty of Physics, University of
Vienna, Boltzmanngasse 5, A-1090 Vienna, Austria
\end{affiliations}

\begin{abstract}
Electrons in isolated graphene layers are a two-dimensional gas of
massless Dirac Fermions. In realistic devices, however, the
electronic properties are modified by elastic deformations,
interlayer coupling and substrate interaction. Here we unravel the
electronic structure of doped graphene, revisiting the stage one
graphite intercalation compound KC$_8$ using angle--resolved
photoemission spectroscopy and ab--initio calculations. The full
experimental dispersion is in excellent agreement to calculations
of doped graphene once electron correlations are included on the
$GW$ level. This highlights that KC$_8$ has negligible interlayer
coupling. Therefore Dirac Fermion behaviour is preserved and we
directly determine the full experimental Dirac cone of doped
graphene. In addition we prove that superconductivity in KC$_8$ is
mediated by electron--phonon coupling to an iTO phonon, yielding a
strong kink in the quasiparticle dispersion at 166~meV. These
results are key for understanding, both, the unique electronic
properties of graphene and superconductivity in KC$_8$.
\end{abstract}
\maketitle

The recent discovery of two--dimensional meta--stable graphene
sheets has sparked enormous interest in their low--energy
electronic
structure~\cite{geim07-review,thomas07-naturematerials,novoselov06-graphite,lanzara06a-graphite}.
Graphene samples are in general obtained by three methods: (1)
repeated peeling of a graphite single
crystal~\cite{novoselov06-graphite}, (2) growth by chemical vapour
deposition on
Ni(111)~\cite{nagashima94-graphene,alex07-graphenenickel} and (3)
precipitating few--layer graphene from SiC~\cite{seyller06-sic}.
Angle--resolved photoemission spectroscopy (ARPES) has been proven
to be a key tool to determine the electronic structure of one and
few layer graphene~\cite{rotenberg06-graphite_bilayer} and
graphite~\cite{lanzara06a-graphite,alex06-correlation,takahashi07-prl}.
A major problem for the investigation of substrate based graphene
is that there is a significant modification of the electronic
structure due to interaction with the substrates yielding charge
transfer and hybridisation~\cite{alex07-graphenenickel}.

One way to overcome this problem is to measure single crystalline
graphite, which has no substrate interaction. In this case
however, a $k_z$ dispersion of two $\pi$ valence
bands~\cite{alex06-correlation} and a small
gap~\cite{orlita08-gap} were observed because of the AB stacking
sequence of graphene layers. Another important issue in both
systems is the renormalization of the bare electronic band
structure due to doping dependent electron--electron
correlation~\cite{alex06-correlation}, electron--phonon
coupling~(EPC)~\cite{takahashi07-prl,rotenberg08-ca} or
electron--plasmon coupling~\cite{rotenberg06-graphite}.

In order to circumvent the problems of substrate interaction,
strong bilayer splitting and electron--electron correlation we
revisited stage one graphite intercalation compound (GIC) KC$_8$.
GICs have been at the focus of intense research in the last four
decades because they have a wide range of tunable electronic
properties~\cite{dresselhaus81}. Especially, for stage~I GICs
superconductivity was observed with transition temperatures $T_C$
ranging from below 1~K for alkali metal intercalation (e.g. 0.55 K
for KC$_8$~\cite{hannay65-gicsuperconductivity}) up to 11.5 K for
CaC$_6$~\cite{csani05-cac6}. In both cases EPC is the
superconducting pairing
mechanism~\cite{sasaki07-superconductivity,mauri-cac6,csani05-cac6}.
In stage I alkali GICs the graphene layers have $AA$ stacking and
only one $\pi$ conduction(valence) band and can thus be considered
as a doped graphene layer sandwiched in between two positively
charged plates (Supplementary Fig.~S1). Thus the low--energy band
structure of, both, stage~I GICs and graphene are described by a
2$\times$2 tight--binding (TB)
Hamiltonian~\cite{blinkowski80-hamiltonian,saito86-gic} resulting
in a linear $\pi$ band dispersion close to the crossing point of
the valence and conduction bands. The electronic band structure of
GICs was also calculated by density functional theory in the local
density approximation (LDA)~\cite{rukola08-kc8lda}.
Experimentally, preliminary studies on ARPES of GICs were
reported~\cite{eberhardt80-lic6,gunasekara-arpes}. Until now the
details in the low energy quasiparticle (QP) dispersion of GICs
regarding the superconducting coupling mechanism are not
identified. Furthermore the issue of whether the charge transfer
to graphite is
complete~\cite{wang91-intercalation,johnson86-kc8,breitholtz07-kc8,zhang87}
or
partial~\cite{oelhafen-alkali,gunasekara-arpes,eberhardt80-lic6}
was never resolved.
\par
In this work we revisited the electronic structure of KC$_8$ GICs
using a combination of ARPES and ab--initio calculations. We proof
a complete charge transfer and find excellent agreement to
ab--initio $GW$ calculations including electron--electron
correlation. This highlight that a rigid band shift model of
graphene is applicable for KC$_8$ and electron--electron
correlation play a crucial role in the QP dispersion. Hence we
unravel for the first time the full experimental Dirac cone of
graphene without modifications by substrate interactions and
directly determine the corresponding momentum dependent Fermi
velocites ($v_F$) in the valence and conduction bands. Thus our
results are key for understanding the unique doping dependent
electronic transport properties of gated graphene layers in
nanoelectronic devices. In addition we show a detailed study of
the direction dependent renormalization of the QP dispersion at
low binding energy that highlights the coupling to an iTO phonon
at 166~meV. We directly show that the coupling to this phonon,
which is also responsible for the double--resonance Raman process,
is the major contribution to the superconducting pairing in
KC$_8$.

\begin{figure}
\includegraphics[width=17cm]{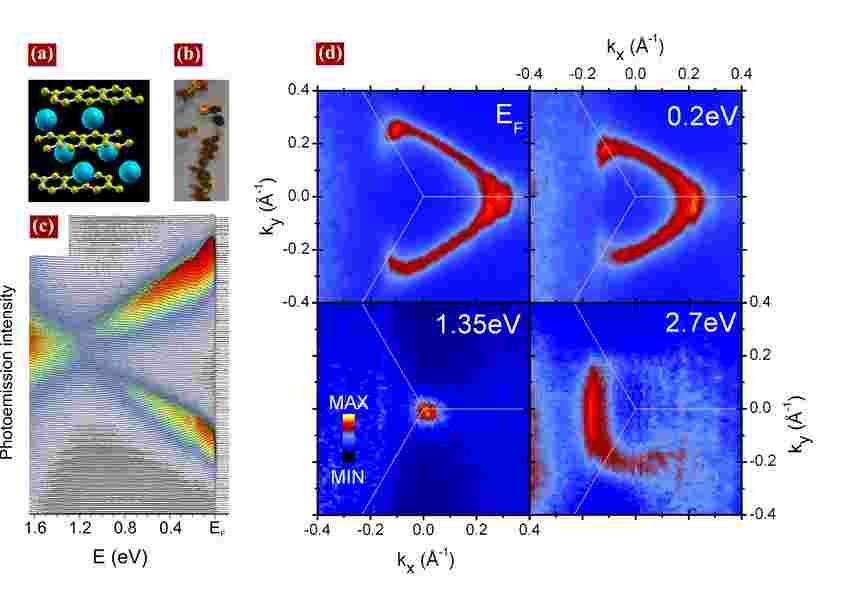}
\caption{(a) Crystal structure of KC$_8$. (b) Photo of intercalated
KC$\rm _8$ single crystals with golden colour. (c) As measured EDC cuts through the
corner of the 3D BZ. (d) Equi--energy contours of the photoemission intensity
taken at 48~eV photon energy for four binding energies.\label{fig:arpes1}}
\end{figure}

\begin{figure}
\includegraphics[width=17cm]{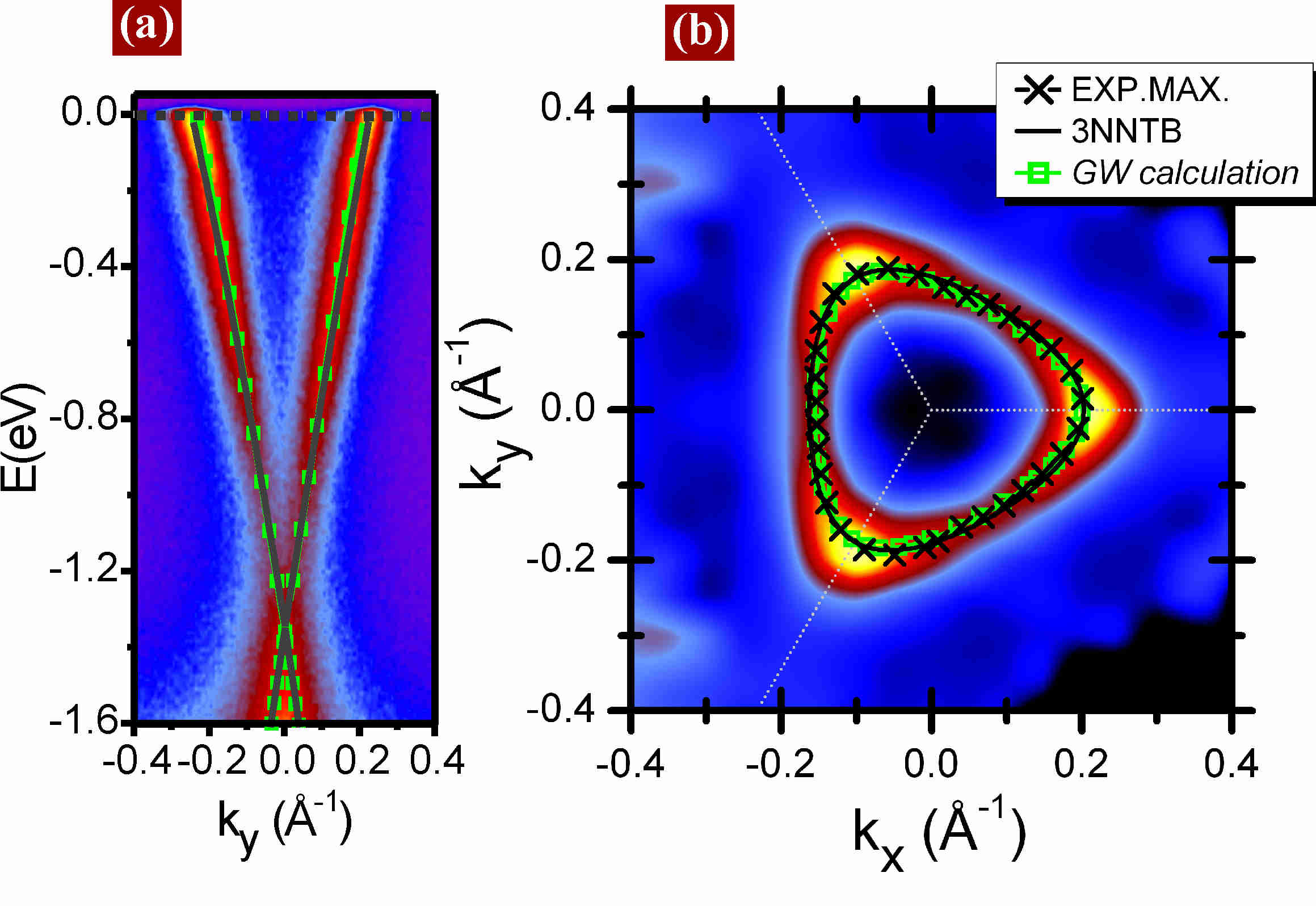}
\caption{(a) ARPES scan measured close to the $k_y$ direction
along with the TB fit of the bare-band dispersion (black) and the
$GW$ calculation for electrostatically doped graphene (blue) and
undoped graphene (green circles). (b) Symmetrized equi--energy
contour for E=0.24~eV and maxima (crosses) along with the TB fit
and the $GW$ ab--initio calculations. \label{fig:theory}}
\end{figure}

The KC$_8$ crystal structure is given by individual graphene
sheets separated by layers of potassium as shown in
Fig.~\ref{fig:arpes1}(a). This compound was synthesized in--situ
by evaporation of potassium as described in detail in the Methods
section. The fully intercalated graphite crystals have a
characteristic golden colour as shown in Fig.~\ref{fig:arpes1}(b).
We now carry out a detailed analysis of the QP dispersion of
KC$_8$ as measured by ARPES. In Fig.~\ref{fig:arpes1}(c) we show
energy dispersion curves (EDC) and cuts of the $\pi$ bands that
intersect the corners of the Brillouin zone (BZ). The cuts are
done close to the $k_y$ direction [see coordinate system in
Fig.~\ref{fig:photohole}(b)]. We also measured the photon energy
dependence and, beside changes in the matrix elements, do not
observe a dispersion perpendicular to the layers.
Fig.~\ref{fig:arpes1}(d) shows equi--energy cuts of the raw
photoemission intensity. When moving from the cut at $E_F$ to
lower energies the circumference of the equi--energy contour of
the conduction band becomes smaller. Evidently we have one $\pi$
valence and one $\pi$ conduction band and they meet in one point.
Therefore in akin analogy to graphene the crossing point at
1.35~eV is from now on called the "Dirac point". Decreasing the
energy of the cut below 1.35~eV enlarges the equi--energy contour
of the valence band. Comparing two contours with the same distance
from the Dirac point [e.g. cuts at $E_F$ and 2.7~eV in
Fig.~\ref{fig:arpes1}(d)], it can be seen that the $\pi$ valence
band is steeper and has less trigonal warping. Such a result is
related to the fact that the self--energy corrections to the
trigonal warping effect increase with doping~\cite{roldan08-ee}.
Interestingly, as depicted in Fig.~\ref{fig:arpes1}(d), the
photoemission matrix elements of the valence and conduction bands
are complementary yielding a high photoemission intensity in 1/3
and 2/3 of the BZ, respectively. From the contour at $E_F$ in
Fig.~\ref{fig:arpes1}(d) we determine the number of carriers $n_e$
by integrating the volume inside the Fermi surface. This yields
$n_e=7.2\times 10^{21}$ electrons cm$^{-3}$, close to a full
charge transfer ($n_e=8.9\times 10^{21}$ electrons cm$^{-3}$). The
remaining deviations can be explained by a slightly lower
stochiometry, i.e. K$_{0.85}$C$_8$.  From these results,
concomitant with the absence of any Fermi surface of K~4$s$ states
close to the $\Gamma$ point (Supplementary Fig.~S7) we can safely
state a complete charge transfer from potassium in agreement with
previous
results~\cite{wang91-intercalation,johnson86-kc8,breitholtz07-kc8,zhang87}.
As can be seen in Fig.~\ref{fig:theory} the ARPES intensity maxima
at $E_F$ and the linear dispersion close to the Dirac point are
very well described by a TB fit including third nearest neighbours
(NN) (Supplementary Table 1). First NN TB calculations are
inadequate and can not describe the experimental equi--energy
contours. This new TB fit reproduces very well the experiments
shown in Fig.~\ref{fig:theory} and will be also used for the
self--energy analysis of the EPC to be described below.
\par
In order to further understand the underlying band structure on an
ab--initio level we compared the experimental results to
calculations at six different levels comprising LDA and
$GW$~\cite{hybert86,hedin65-gw,louie-gw} calculations of (un)doped
graphene and KC$_8$. As seen in the Supplementary in Fig.~S2-S6,
band structures calculated at the level of LDA are too flat
yielding an incomplete charge transfer with a partially occupied
K~4$s$ band in the vicinity of the $\Gamma$ point. Calculations on
the $GW$ level including full and partial self consistency in G
are in much better agreement to the experiment. The obtained
results for the QP dispersion at $K$ are identical but only the
latter explains the lack of photoemission intensity at the
$\Gamma$ point in KC$_8$ (Supplementary Figs.~S7-S8). As depicted
in Fig.~\ref{fig:theory}(a) the observed linear dispersion is
reminiscent of the underlying structure of the graphene parent
compound and in perfect agreement to the $GW$ calculations of
graphene. This also holds for the equi--energy contour  of the
photoemission intensity at $E_F$ shown in
Fig.~\ref{fig:theory}(b). These results clearly indicate that
electron--electron correlation is crucial to explain the size of
the Fermi surface and the observed trigonal warping. I.e.
electron--electron correlation is at the heart of the band
structure the underlying (doped) graphene layers. Hence our
results unambiguously highlight the following facts inherent to
the electronic structure of doped graphene without substrate
interaction: (i) the shape of the Dirac Fermions depends on the
doping level and the equi--energy contours above and below the
Dirac point are anisotropic; (ii) the trigonal warping increases
as doping increases; (iii) in KC$_8$ each potassium atom transfers
one electron to the graphene layers.

\begin{figure}
\includegraphics[width=17cm]{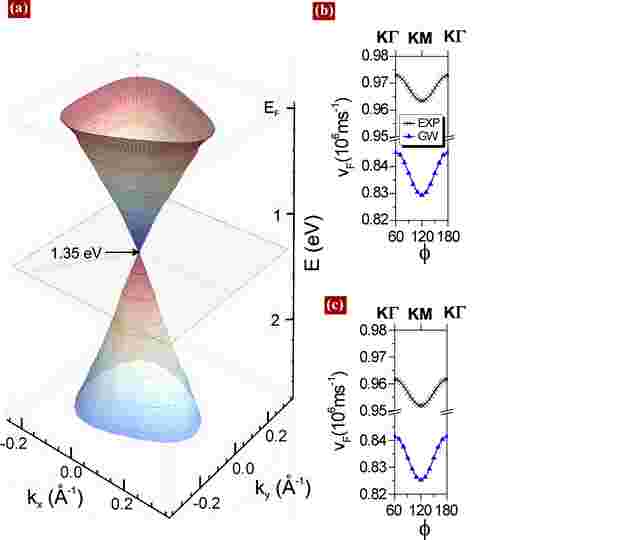}
\caption{(a) Experimental Dirac cone from the observed
photoemission intensity maxima (denoted as dots). Measured and
calculated ($GW$) values of $v_F$ for (b) electrons and (c) holes
around the Dirac point.\label{fig:barrel}}
\end{figure}

\begin{figure}
\includegraphics[width=17cm]{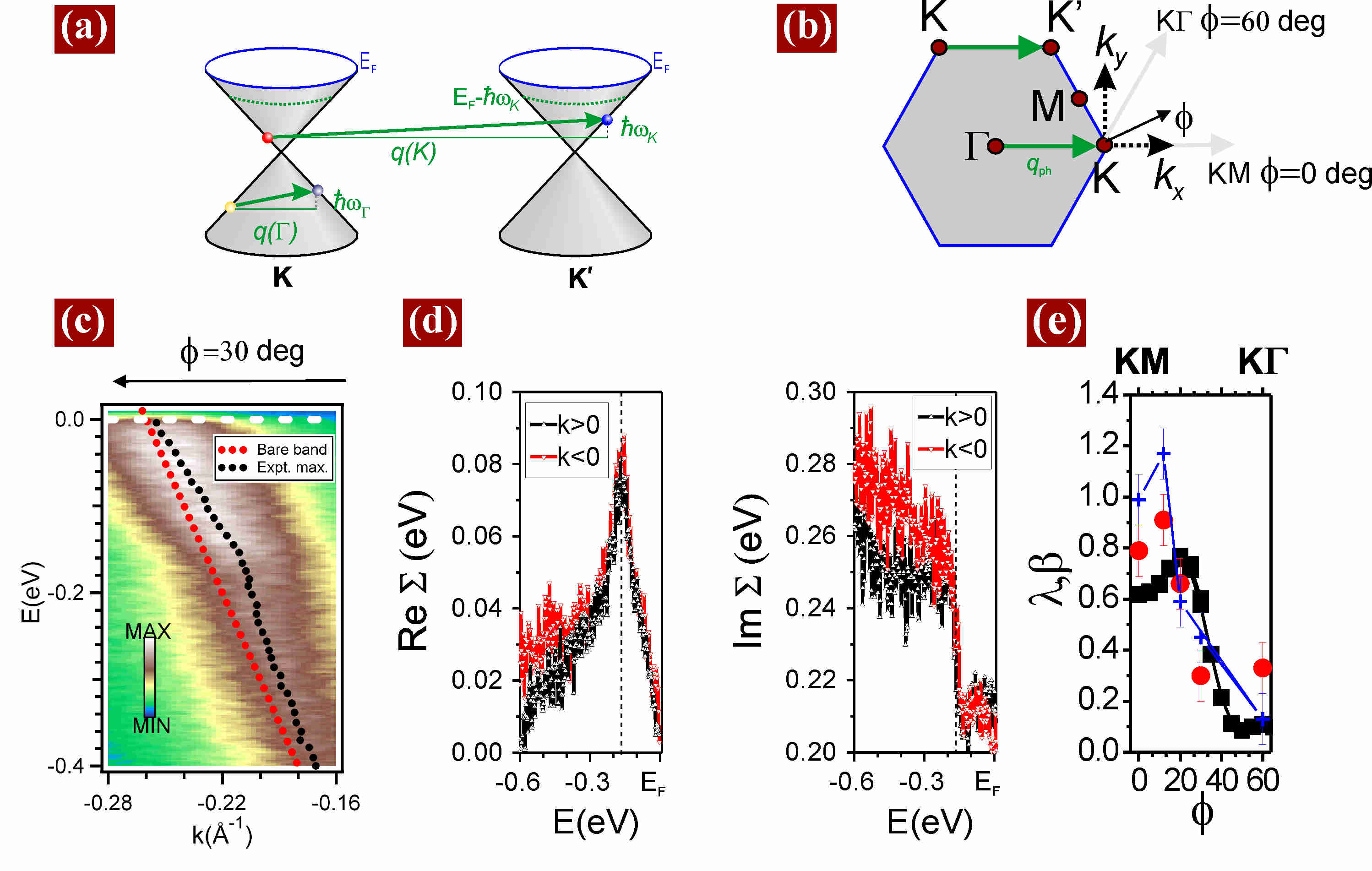}
\caption{(a) Schematics of the photohole scattering process from
$K$ to $K'$ by EPC to a $K$ point phonon with energy
$\hbar\omega_{ph}(q_{ph})$. The initial and final photohole states
around $K$ and $K'$ are denoted by circles. The arrow denotes the
exchanged phonon with wavevector $q_{ph}$ and energy
$\hbar\omega_{ph}(q_{ph})$ which is emitted by photohole
relaxation. (b) 2D BZ indicating the phonon wavevector $q_{ph}$.
(c) measured ARPES intensity in the region of the kink. A straight
dashed line denotes the bare band dispersion and a kinked dotted
line are the experimental MDC maxima. (d) real and (e) imaginary
part of the self--energy. (f) the angular dependence of the EPC
constant $\lambda$ and the electron--electron correlation constant
$\beta$; black squares(red circles) denote $\lambda$ extracted
from the measured real(imaginary) part of the self--energy.
$\beta$ is denoted by blue crosses. \label{fig:photohole}}
\end{figure}

Most importantly let us now turn to a quantitative assessment of
the QP bandstructure around the Dirac point, i.e. unravelling for
the first time the full experimental Dirac cone. For this purpose,
we extracted the photoemission maxima for binding energies between
$E_F$ and 3~eV in steps of 10~meV. The results are depicted in
Fig.~\ref{fig:barrel}(a). It is obvious that the two--dimensional
band structure close to the Dirac point is linear and two bands
meet each other in one point. This highlights that KC$_8$ indeed
consists of doped sheets of graphene and can be well described
with the aforementioned TB and $GW$ approaches. We point out this
this is in contrast to all previous experiments on substrate based
graphene, clearly observing the opening of a gap in the electronic
and structure of epitaxial graphene on
Ni(111)~\cite{alex07-graphenenickel} and
SiC~\cite{lanzara07-graphene}. Although the origin of the gap in
graphene on SiC is under debate~\cite{lanzara07-comment} it
existence is accepted. The opening of a gap causes a breakdown of
Dirac Fermions which has profound limitation on the observation of
relativistic physics in graphene. Therefore the spectroscopic
investigation of doped graphene layers in KC$_8$ provides an
elegant solution to this problem.

Hence, by analyzing the full two-dimensional dispersion of KC$_8$
around the Dirac point we can thus actually learn about the
physics of graphene. In order to further underline the validity of
this analogy we analyze the experimental direction dependent $v_F$
in the region of 0.1~eV above and below the Dirac point. These
results for the valence and conduction band along with the $GW$
calculations of doped graphene are depicted in
Fig.~\ref{fig:barrel}(b) and Fig.~\ref{fig:barrel}(c),
respectively. The $v_F$ has a maximum(mininimum) in
$K\Gamma$($KM$) direction. The very good quantitative agreement
highlights the validity of our approach. In close similarity to
our results on pristine graphite~\cite{alex06-correlation}, we
observed that electron--electron correlation is crucial to explain
the band structure of doped graphene layers in KC$_8$.

We now turn to the analysis of QP bandstructure of KC$_8$ close to
E$_F$ and their implications for superconductivity. In addition to
the linear dispersion, as can be see in Fig.~\ref{fig:arpes1}(c)
Fig.~\ref{fig:theory}(a) and Fig.~\ref{fig:barrel}(a), the $\pi$
conduction band of KC$_8$ is strongly kinked at 166~meV. This is
in agreement to previous results on doped
graphene~\cite{rotenberg08-ca}. By an accurate comparision to the
graphite phonon dispersion
relation~\cite{j920,ludger04-phonon,maultzsch04} we can
unambiguously assign the kink to a coupling to the in--plane
transverse optical (iTO) phonon branch near the $K$ point (see the
Supplementary Table~2 for the calculated phonon frequencies
including non-adiabatic effects). This agrees perfectly with both
the energy of the kink and also with the facts that the EPC matrix
element with the iTO phonons is much stronger than with $\Gamma$
point phonons~\cite{basko08-elph} and that the phonon density of
states is strongly peaked due to the flat dispersion of the iTO
branch around $K$. The photohole decay process is shown
schematically in Fig.~\ref{fig:photohole}(a). It is clear that a
photohole can relax to a lower binding energy state in two ways by
intravalley and intervalley scattering. For the case of
intravalley scattering the photohole scatters around $K$ or $K'$
points and relaxes by emission of $\Gamma$ point phonons. However,
as their energy does not fit with the measured kink we disregard
this mechanism. Therefore the relevant process is related to
intervalley scattering where the photohole scatters between $K$
and $K'$ points and relaxes by emitting phonons close to the $K$
point. In Fig.~\ref{fig:photohole}(b) we illustrate this mechanism
in the 2D BZ of graphene with the exchange of a phonon with
wavevector $q_{ph}$. Interestingly, a very similar process for
scattering of photoexcited electrons is responsible for the $D$
and $G'$ band in the double resonance Raman process in $sp^2$
hybridized carbon materials.

Once we have identified the phonon mode responsible for the
observed kink it is important to accurately determine $\lambda$,
the EPC constant~\cite{fink04-reevaluation}. In recent
works~\cite{rotenberg08-ca,valla08-cac6} $\lambda$ was estimated
from the change in the slope of linear bare electronic bands;
however this is not justified as the proper bare band structure at
the $E_F$ has to be considered, as was clearly pointed out in a
recent theoretical work~\cite{louie-electronphonon}. In the
present work we avoid this problem by considering the total
self--energy $\Sigma(E)$ from which $\lambda$ can be directly
obtained by a simple energy derivative at $E_F$
$\lambda=-\frac{\partial Re(\Sigma(E))}{\partial E}|_{E=E_F}$. To
do so we fit the momentum distribution curves (MDCs) of the QP
band structure by Lorentzians. The positions of the maxima and the
widths of the MDCs are directly proportional to the real and
imaginary part of $\Sigma$, i.e. Re($\Sigma$) and Im($\Sigma$),
respectively~(see Supplementary Figs.~9-10). The QP dispersion in
the energy region close to the kink is shown in detail in
Fig.~\ref{fig:photohole}(c) along with the measured maxima of MDCs
and the bare band dispersion (the cut is in $k_y$ direction in
between $KM$ and $K\Gamma$ and thus corresponds to an averaged
$\lambda$). In the simple case of coupling to one Einstein phonon
Re($\Sigma$) has a peak and Im($\Sigma$) has a jump at the energy
of the coupling phonon mode~\cite{fink04-reevaluation}. Indeed
this is what we observe in the Figs.~\ref{fig:photohole}(d,e) due
to the strong coupling to the iTO phonon from $K$ point at 166meV.
Here we emphasize that the evaluation of $\lambda$ from
Im($\Sigma$) does not depend on the details of the dispersion of
the bare bands. For the analysis of $\lambda$ from Re($\Sigma$) we
used the TB calculation of the bare bands which gives the correct
curvature in $KM$ and $K\Gamma$ directions. Thus our analysis does
not suffer from an overestimation of $\lambda$ in $KM$ direction.
Moreover, from the slope of Re($\Sigma$) and from the height of
the jump in Im($\Sigma$) it is possible to evaluate the EPC
$\lambda$\cite{fink04-reevaluation}: the values we got are
$\lambda=0.4$ and $\lambda=0.3$ from the analysis of Re($\Sigma$)
and Im($\Sigma$), respectively. This procedure can be used to
obtain the wavevector dependence of $\lambda_k$. In
Fig.~\ref{fig:photohole}(f) we show that the EPC obtained from
both the Re($\Sigma$) (squares) and Im($\Sigma$) (circles): we
thus obtain that $\lambda_k$ is maximum close to $KM$ direction
and minimum in the $K\Gamma$ direction. As can be seen in the
Fig.~\ref{fig:photohole}(f) the resulting $\lambda$ from
Re($\Sigma$) and Im($\Sigma$) are in good agreement regarding the
anisotropy of the coupling although the actual size of the
coupling differs by about 20 \%. The minor anisotropy obtained in
previous DFT calculations~\cite{louie-electronphonon} compared to
the one extracted from the present work
[Fig,~\ref{fig:photohole}(f)] and other experimental
works~\cite{rotenberg08-ca,valla08-cac6} raises some concerns
related to the need of going beyond LDA to describe exchange and
correlation effects in the QP energies and wavefunctions. As we
stated above, self-energy corrections at the $GW$ level are needed
to describe the observed absence of ARPES intensity at $\Gamma$,
in contrast to what is predicted by LDA. Another possible
explanation to partly explain the strong anisotropy in $\lambda_k$
observed experimentally might be the phonon trigonal warping
effect~\cite{q953}. Thus further work is needed to address the
phonon dispersion relation and the impact of the $GW$ self--energy
corrections on EPC~\cite{basko08-elph} which we are currently
working on. We have also analyzed $\beta$, i.e. the coefficient of
the quadratic energy dependence of Im($\Sigma$) which is depicted
as crosses in Fig.~\ref{fig:photohole}(f) (see also Supplement
Fig.~10). Interestingly, $\lambda$ and $\beta$ have the same
anisotropy, which reflects the anisotropy of the QP bandstructure
at $E_F$.

The strong EPC limits the maximum bias current through graphene
based nanoelectronic devices and is responsible for the strong
double--resonance Raman scattering in $\rm sp^2$ bonded carbon
materials. In classical BCS theory, with a pairing mechanism based
on EPC it is also one of the limiting factors for the transition
temperature of KC$_{\rm 8}$. The evaluated transition temperature
using the McMillan formula~\cite{mcmillan-tc} with the high phonon
temperature $\Theta=1926$~K and the average EPC constant of
$\lambda=0.45$ [Fig.~\ref{fig:photohole}(f)] and a screened
pseudopotential $\mu^\star\sim 0.14$~\cite{mauri-cac6} would allow
a $T_C$ up to 6~K. These remaining differences to the experimental
$T_C$ can be explained by either uncertainties in $\mu^\star$ or
by additional not yet known pair breaking effects.
\par

In conclusion we have synthesized $ \rm KC_8$ in--situ and
performed ARPES measurements. We have found a complete charge
transfer and that the QP dispersion of KC$_8$ is strongly modified
in the presence of EPC to the iTO phonon branch at $K$. The value
of the coupling constant is in good agreement to the experimental
value of $T_C$ and hence determines the superconducting pairing
mechanism. Interestingly, this iTO phonon branch is also
responsible for the strong double resonance Raman scattering in
$\rm sp^2$ bonded graphite materials. The QP band structure of
KC$_8$ is in good agreement to ab-initio calculations for graphene
after including self--energy corrections on a $GW$ level. We thus
conclude that the conical band dispersion of $\rm KC_8$ closely
resembles that of Dirac Fermions of graphene after applying a gate
voltage. Our results circumvent the problems which are associated
with strong substrate interaction of
graphene~\cite{lanzara07-graphene,lanzara07-comment,alex07-graphenenickel}
or with interlayer interaction~\cite{alex06-correlation}, both of
which cause a breakdown of Dirac Fermions. Most importantly, we
unravelled for the first time the full experimental Dirac cone of
doped graphene and directly determine the corresponding momentum
dependent $v_F$ in the valence and conduction band as such
providing crucial input for the understanding of the unique
electronic and transport properties of graphene.

\begin{methods}
Experiments were done at BESSY~II using the UE112-PGM2 beamline
and a Scienta RS~4000 analyzer yielding a total energy resolution
of 15~meV and a momentum resolution better than $\rm
0.01~\AA^{-1}$. Natural graphite single crystal samples were
mounted on a three axis manipulator and cleaved in--situ.
Intercalation was performed with the sample at room temperature by
evaporating potassium metal from a commercially available SAES
getter source. In order to check the doping level, we measured
ARPES after each intercalation step. The proof that we reached
stage~I was given by the appearance of only one $\pi$ valence band
[instead of 2(3) valence bands for stages II(III)]. In addition
stage~I compounds are identified by their characteristic golden
color. After full intercalation, KC$_8$ was immediately cooled
down by liquid He to 25~K and measured. The calculations of the
electronic dispersion of graphene in a slab geometry ($d=20$~a.u.)
are performed on two levels. First, we calculate the Kohn-Sham
band-structure within the LDA to density-functional
theory~\cite{abinit}. Wave functions are expanded in plane waves
with an energy cutoff at 25 Ha. Core electrons are accounted for
by Trouiller-Martins pseudopotentials. In the second step, we use
the $GW$ approximation \cite{hybert86,hedin65-gw,louie-gw} to
calculate the self-energy corrections to the LDA
dispersion~\cite{self}. For the calculation of the dielectric
function $\epsilon(\omega,q)$ we use a Monkhorst-Pack $k$ grid
sampling 36$\times$36$\times$1 points and bands up to 70eV (namely
50 bands) of the first BZ. In a further step we performed $GW$
calculations including partial self consistency in $G$. Note, that
this is a conserving approximation whereas the single shot
G$_0$W$_0$ is not.
\end{methods}

\begin{addendum}
\item A.G. acknowledges a Marie Curie Individual Fellowship (COMTRANS)
from the European Union. T.P. acknowledges DFG projects PI 440/3
and 440/4. C.A. and A.R. acknowledge funding by the Spanish MEC (FIS2007-65702-C02-01),
"Grupos Consolidados UPV/EHU del Gobierno Vasco" (IT-319-07), and by the
European Community through NoE Nanoquanta (NMP4-CT-2004-500198), e-I3
ETSF project (INFRA-2007-1.2.2: Grant Agreement Number 211956)
SANES(NMP4-CT-2006-017310), DNA-NANODEVICES (IST-2006-029192) and
NANO-ERA Chemistry projects and the computer resources provided by
the Barcelona Supercomputing Center, the Basque
Country University UPV/EHU (SGIker Arina).

 \item[Competing Interests] The authors declare that they have no
competing financial interests.
 \item[Correspondence] Correspondence and requests for materials
should be addressed to A. G. \\(email: ag3@biela.ifw-dresden.de).
\end{addendum}

\end{document}